\documentclass[aps,prx, amsmath,amssymb,
preprint,groupedaddress]{revtex4-2}

\usepackage{graphicx}
\usepackage{dcolumn}
\usepackage{bm}

\usepackage{mathtools}
\usepackage{lipsum}

\newcommand{\GS}{G_\mathrm{signal}}
\newcommand{\GN}{G_\mathrm{noise}}

\newcommand{\R}{\bm{R}}
\newcommand{\F}{\bm{F}}
\newcommand{\YBM}{Y_\mathrm{BM}}
\newcommand{\VBM}{V_\mathrm{BM}}
\newcommand{\VCP}{V_\mathrm{CP}}
\newcommand{\Vtop}{V_\mathrm{top}}
\newcommand{\Vint}{V_\mathrm{int}}

\newcommand{\YCP}{Y_\mathrm{CP}}

\newcommand{\Pb}{\bar{P}}

\newcommand{\dx}[1]{\frac{d#1}{dx}}

\newcommand{\dxii}[1]{\frac{d^2#1}{d\chi^2}}

\newcommand{\GD}{G_\mathrm{1D}}
\newcommand{\GDb}{\bar{G}_\mathrm{2D}}
\newcommand{\GDD}{\hat{G}_\mathrm{2D}}
\newcommand{\GDN}{\tilde{G}_\mathrm{2D}}

\newcommand{\GDDD}{{G}_\mathrm{2D}}
\newcommand{\GTD}{{G}_\mathrm{3D}}

\newcommand{\Nrms}{N_\mathrm{rms}}
\begin{document}


\title{The Noise Within: Signal-to-Noise Enhancement \\[2pt]
  via Coherent Wave Amplification in the Mammalian Cochlea}
\author{Alessandro Alto\`e}
 \email{altoe@usc.edu}

\author{Christopher A.\ Shera} 
\altaffiliation[Also at ]{Department of
  Physics \& Astronomy, University of Southern California; christopher.shera@usc.edu}
\affiliation{Auditory Research Center\\ Caruso Department of
  Otolaryngology\\ University of Southern California\\ Los Angeles, CA
  90033
}%
\begin{abstract}
\noindent
The extraordinary sensitivity of the mammalian inner ear has
captivated scientists for decades, largely due to the crucial role
played by the outer hair cells (OHCs) and their unique electromotile
properties. Typically arranged in three rows along the sensory
epithelium, the OHCs work in concert via mechanisms collectively
referred to as the ``cochlear amplifier'' to boost the cochlear
response to faint sounds.  While simplistic views attribute this
enhancement solely to the OHC-based increase in cochlear gain, the
inevitable presence of internal noise requires a more rigorous
analysis. Achieving a genuine boost in sensitivity through
amplification requires that signals be amplified more than internal
noise, and this requirement presents the cochlea with an intriguing
challenge.  Here, we analyze the effects of spatially distributed
cochlear-like amplification on both signals and internal noise.  By
combining a straightforward but powerful mathematical analysis with a
simplified model of cochlear mechanics designed to capture the
essential physics, we generalize previous results about the impact of
spatially coherent amplification on signal degradation in active gain
media. We identify and describe the strategy employed by the cochlea
to amplify signals more than internal noise and thereby enhance the
sensitivity of hearing.  For narrowband signals, this elegant,
wave-based strategy consists of spatially amplifying the signal within
a localized cochlear region, followed by rapid attenuation.
Location-dependent wave amplification and attenuation meet the
necessary conditions for amplifying near-characteristic frequency (CF)
signals more than internal noise components of the same frequency.
Our analysis reveals that the sharp wave cut-off past the CF location
greatly reduces noise contamination.  The distinctive asymmetric shape
of the ``cochlear filters'' thus underlies a crucial but previously
unrecognized mechanism of cochlear noise reduction.

\end{abstract}
  
\maketitle
  
\section{Introduction}
In the 19th century, Bernhard Riemann made the remarkable observation
that the sound of a foghorn could be heard from a distance of five
miles. He concluded that the human ear must be capable of detecting
sounds that generate only sub-atomic motions of the eardrum
\cite{bell2022}. During the succeeding one and a half centuries,
Riemann's conjecture has been repeatedly verified
\cite{dalhoff2007}. The extraordinary sensitivity of the mammalian ear
can be attributed to the coordinated, piezoelectric behavior of outer
hair cells (OHCs) \cite{brownell1985}. Arranged in
rows along the sensory tissue (the organ of Corti), these cells act as
actuators capable of boosting sound-induced vibrations of the sensory
tissue by more than two orders of magnitude \cite{fisher2012}. The
prevailing belief in the field posits that OHCs actively amplify
sound-induced waves as they propagate along the spiral structure of
the cochlea. Collectively, the mechanisms involved are
known as the ``cochlear amplifier''.

However, whether cochlear amplification constitutes a viable strategy
for enhancing the sensitivity of hearing remains controversial.
Because the minimum signal level to which sensory neurons can
meaningfully respond is inherently limited by the level of internal
noise \citep[see e.g.,][]{vannetten2003}, it remains unclear how the
cochlear amplifier, while amplifying signals, can avoid amplifying the
accompanying internal noise \cite{vanderheijden2021}.  Although the
dominant sources of intracochlear noise remain to be firmly
identified---these necessarily include both thermal noise and noise
associated with the stochastic gating of the hair-cell ion channels
\citep[see e.g.,][]{vannetten2003,sasmal2018}---intrinsic cochlear
mechanical noise is both present and measurable, and it depends on the
same mechanisms that control signal amplification
\cite{nuttall1997}. While previous work has focused on the effects of
internal noise sources on the sensitivity of hair-cell stereocilia
\citep[see e.g.,][]{vannetten2003,sasmal2018}, the amplification of
intracochlear noise remains unexplored.

In this study, we investigate the impact of spatially distributed
amplification on both signals and internal noise using two distinct
but complementary approaches: a mathematical model of spatially
distributed amplification and an active model of the cochlea. We begin
by examining the simplest scenario, which involves a highly
anisotropic, one-dimensional medium comprising a series of cascaded
``noisy'' amplifiers in which signals and noise propagate in only a
single direction. We then move to the more challenging but
biologically relevant case where the medium is nearly isotropic, so
that signals and noise propagate and are amplfied in both directions.
Finally, we investigate signal and noise amplification within a
simplified but physically realistic linear model of the
cochlea. Importantly, our analysis concerns only noise sources that
are located within the cochlea: the ear processes external noise in
the same way that it processes signals \cite{recio2005}. Furthermore,
as we all know from cocktail parties or an old-fashioned bar fight
over a jukebox, which sounds are ``signals'' and which are ``noise''
depends entirely on what one wants to listen to.


\section{Spatially distributed amplification in noisy active media}
\begin{figure*}[hbp]
\begin{center}
\includegraphics{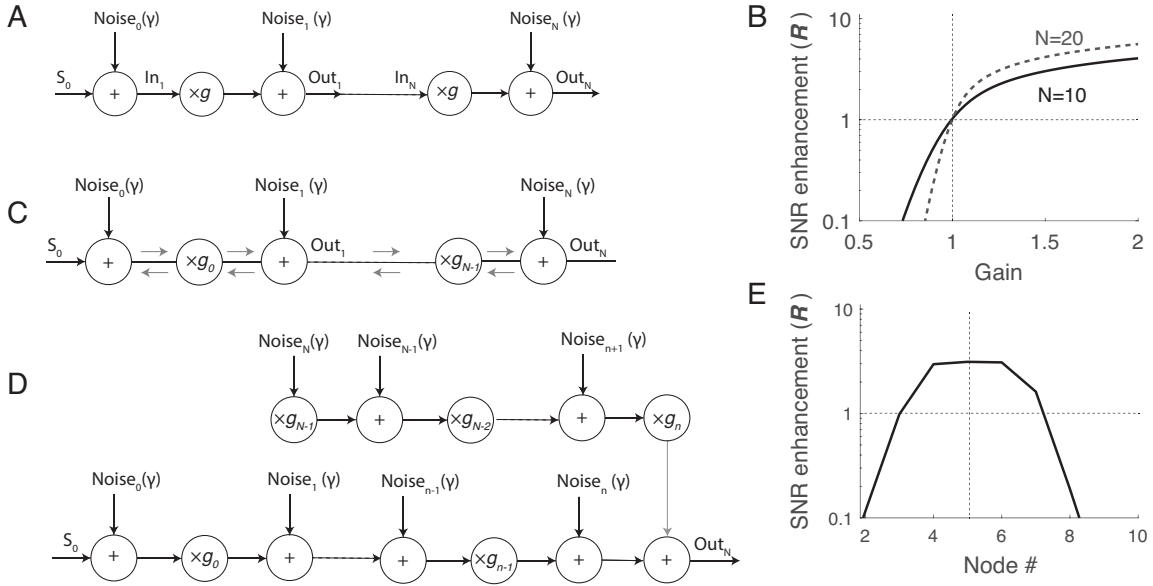}
\caption{ A) Effect of spatially distributed ``one-way'' amplification
  on signal and internal noise. The model consists of a chain of
  linear amplifiers (multipliers) with gain $g$; the effect of
  internal noise is simulated by adding noise before and after each
  amplification stage. B) SNR enhancement (${\bf R}$) at the $N$-th
  node of the amplifier chain (shown for $N=10$ and $N=20$) as a
  function of the amplifier gain, $g$. C) Bidirectional noisy
  amplification model. In this model, internal noise propagates
  and is amplified identically in both directions. D) Equivalent one-way
  amplification model to the study noise and signal response at the $n$-th
  node. E) Example of enhancement factor at different nodes in a chain
  of $N=10$ bidirectional amplifiers. In this example, the amplifier
  gain is chosen to improve the SNR at node 5 (see text) by setting
  $g_m=3$ for $m<5$ and $g_m=0.1$ for $m\ge5$.}
\label{fig:1}
\end{center}
\end{figure*}

{\em Propagation of signals and noise in one direction.}  We start by
considering the simple scenario of the distributed ``one-way'' noisy
amplifier, depicted in Fig.\,\ref{fig:1}A. The model consists of a
chain of amplifiers that multiply the input signal ($S[0]$) by a
factor $g$, representing the amplifier gain. The medium's noise is
represented by noise sources that are summed with the propagating
signal after each amplification stage. To remove the ambiguity
regarding whether noise should be included before or after the
amplification stage, the model includes noise sources located both at
the input of the first amplifier and at the output of last. This model
approximates a strongly anisotropic medium, where signals and noise
propagate only in one direction (from left to right in
Fig.\,\ref{fig:1}A). This scenario accurately represents what occurs
in many man-made systems, such as cascaded electronic amplifiers or
radio repeaters---indeed the formula we derive here are essentially
the same used to calculate the noise figure of cascaded electronic
amplifiers \cite{pettai}.

In this model we can turn amplification ``off''---and thereby model
signal propagation in a lossless, noisy medium---by imposing the
condition $g=1$. Or we can turn it ``on'' by setting $g\neq1$.  When
$g>1$, the chain amplifies signals as they propagate.  When $g<1$, the
distributed amplifiers become distributed, attenuating ``brakes.''  By
comparing signal and noise for the three conditions ($g=1$, $g>1$, and
$g<1$), we quantify the impact of amplification and attenuation on the
SNR along the chain (i.e., at the nodes out$_\mathrm{1,2...n}$ in
Fig.\,\ref{fig:1}A).

The root-mean-square (rms) amplitude of the signal at a given node $n$
is simply the rms amplitude of the input signal passed through $n$
multipliers ($S_\mathrm{rms}[n]=g^n S_\mathrm{rms}[0]$).  Turning on
the amplifier thus boosts the signal amplitude by the factor
\begin{equation}
  \GS[n]=g^n\;.
  \label{eq:signal_gain1}
\end{equation}

We focus our analysis on the physically relevant case where the
noise sources are uncorrelated, meaning that the noise in the medium
is spatially incoherent.  For simplicity, we assume that the various
noise sources are independent versions of the same stochastic process,
with rms amplitude $\gamma$.  In this case, the rms amplitude of the
noise ($N_\mathrm{rms}$) at node $n$ can be calculated by incoherent
summation (i.e., linear summation of power) of the various amplified noise
terms. Specifically, the noise power at node $n$ can be expressed as a
geometric series, where the $m$-th term represents the contribution of
the $(n-m)$-th source, amplified (or attenuated) $m$ times. The
expression for $N_\mathrm{rms}[n]$ can be simplified based on
different scenarios:
\begin{eqnarray}
\Nrms[n]=\sqrt{\sum_{m=0}^ng^{2m}}\gamma=
  \begin{cases}\displaystyle{\sqrt{\frac{g^{2(n+1)}-1}{g^2-1}}\gamma} & \text{for } g\neq1, \label{eq:Nrms}\\
\displaystyle{\vphantom{\Biggl(}\sqrt{n+1}\gamma} &  \text{for } g=1.
 \end{cases}
\end{eqnarray}
Hence, turning on the amplifier boosts the noise gain by a factor of
\begin{equation}
\GN[n]=\frac{\Nrms[n]|_{g\neq1}}{\Nrms[n]|_{g=1}}=\sqrt{\frac{g^{2(n+1)}-1}{(n+1)(g^2-1)}}\;. \label{eq:noise_gain}
\end{equation}
The SNR at node $n$ is given by $R_n=S_\mathrm{rms}[n]/\Nrms[n]$. The effect of
amplification on the system's sensitivity can be quantified by the SNR
enhancement factor \citep{loudon1993}:
\begin{equation}
\R[n]=R_n(\mathrm{on})/R_n({\mathrm{off}})=\GS/\GN\;, \label{eq:noise_en}
\end{equation} 
where $R(\mathrm{on})$ and $R(\mathrm{off})$ are the SNR with the
amplifier on ($g\neq1$) and off ($g=1$), respectively.  Figure
\ref{fig:1}B illustrates the enhancement factor as a function of $g$
for two values of $n$.  When $\R>1$ the signal is amplified more than
the internal noise, and the SNR increases at the considered
node. Conversely, when $\R<1$, the signal is amplified less than the
noise, and the SNR decreases. It follows from
Eqs.\,(\ref{eq:signal_gain1},\ref{eq:noise_gain}) that amplification
($g>1$) boosts signals more than internal noise, increasing the SNR at
all nodes. In particular, the larger the gain, the larger $\R$,
resulting in a greater improvement in SNR at any node. Additionally,
the longer the chain of amplifiers, the larger the benefit of
distributed amplification on the SNR and the greater the increase in
the system's sensitivity. Conversely, when the amplifiers act as
attenuators ($g<1$), $\R<1$, meaning that the signal is attenuated
more than the internal noise.

As the signal propagates along the line, noise from the growing
number of contributing sources accumulates.
A relevant measure of the resulting signal degradation 
is the noise
factor $\F_n=R_n/R_0$, which quantifies how the SNR degrades along the
transmission line. In our case
\begin{equation}
\F_n=\sqrt{\frac{g^{2(n+1)}(1-g^{-2})}{g^{2(n+1)}-1}}\;,
\end{equation}
which approaches 1 (i.e., no significant SNR degradation along the
line) when $g\gg1$. Importantly, this result---namely that distributed
amplification prevents signal degradation---generalizes to the case when
internal noise sources are spatially coherent \footnote{The reader can
  easily verify this statement by repeating the analysis presented above with
  one small variation: when noise sources are coherent
  $\Nrms[n]=\sum_{m=0}^ng^{m}\gamma$.}.

{\em Signal vs.\,noise amplification in isotropic active media.}  We
now extend the simple chain-of-amplifiers model described above by
considering the case of an active medium where waves propagate in both
directions, as in the mammalian cochlea \cite{kemp1978}. In our
simplified treatment, we assume that the medium is isotropic. Thus, we
assume that the amplifiers boost signals propagating in either
direction by the same amount (Fig.\,\ref{fig:1}C). We simplify the
analysis further by ignoring potential scattering effects within
the medium and by assuming that the various noise sources all have
equal amplitudes. In this case, however, we allow the amplifier gain to vary
along the line. When considering signal and noise propagation to node
$n$, the system can be depicted as the combination of two ``one-way''
amplification models (Fig.\,\ref{fig:1}D), representing the
contribution from sources located to the right and to the left of the
node $n$. Note that whereas signals come only from the left, noise
comes from both directions.

Signal propagation from a source node $n'$ to a
receiver node $n$ is encapsulated by the discrete Green's function
$G[n,n']$.  In the simplified model, where each node $n$ amplifies the
signal by the factor $g_n$:
\begin{equation}
G[n,n']=\prod_{m=\min(n,n')}^{\max(n,n')-1}g_m\;.  
\end{equation}  
Note that the Green's function is symmetric: $G[n',n]=G[n,n']$.
In this model, the signal is effectively a source at node $0$; 
its amplitude at node $n$ is therefore
\begin{equation}
S_\mathrm{rms}[n]=S_\mathrm{rms}[0]G[n,0]\;.
\end{equation}
The noise response at node $n$ can be decomposed into the incoherent
summation of noise from both the left and right sides of the node
(Fig.\,\ref{fig:1}D):
\begin{align}
  \Nrms[n] = &\sqrt{\sum_{n'=0}^N(G[n,n']\gamma)^2}  \\
  =& \gamma \sqrt{\sum_{n'=0}^n(G[n,n'])^2+\sum_{n'=n+1}^N(G[n,n'])^2}\;.
  \nonumber
\end{align}

In this case, unlike the simpler anisotropic model of Fig.\,\ref{fig:1}A,
amplification is not necessarily beneficial
for the SNR. 
When the goal is to maximize the SNR at node $n$,
the optimal gain distribution along the amplifier chain is
\begin{align}
g_{n'}\gg1 &\; \text{for} \;  n'<n \label{eq:optimal_bi}\\
g_{n'}\ll1 &\; \text{for} \;  n'\ge n\;. \nonumber
\end{align}
In this case, the system approaches the performance of the one-way
amplification model at the $n$-th node. Unlike the one-way
model, however, it is not possible to increase the SNR at all
nodes simultaneously (see Fig.\,\ref{fig:1}E).


\section{Signal vs.\ noise amplification in the mammalian cochlea}
\begin{figure*}[htbp]
\begin{center}
\includegraphics[]{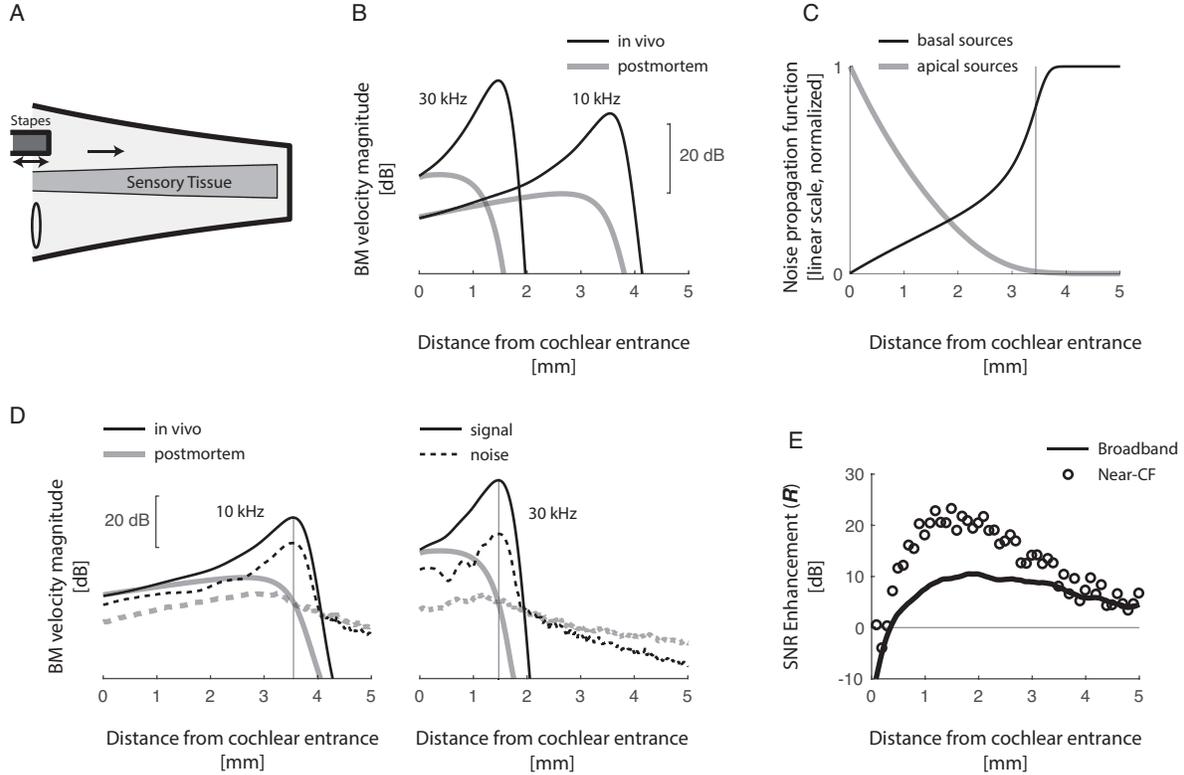}
\caption{A) Simplified anatomical view of the mammalian cochlea. B) BM
  magnitude responses in vivo (amplifier on) and post-mortem
  (amplifier off) to stimulus tones of 10 kHz and 30 kHz calculated in
  a 2D finite-difference model of the mouse cochlea. C) Apical and
  basal noise propagation functions for narrowband noise centered
  around 10 kHz. At each location, these functions quantify the
  expected noise power due to distributed basal and apical noise
  sources of equal strength, respectively. The grey vertical line
  marks the characteristic place. D) BM response magnitude to sound
  signal and narrowband internal noise at 10 and 30 kHz, for both
  postmortem and in vivo models. The curves are normalized so that the
  signal and noise magnitudes at the characteristic places (vertical
  grey lines) are the same postmortem. The difference between in vivo
  signal and noise responses demonstrates that turning on the
  amplifier boosts the SNR at the characteristic place. E) Enhancement
  factor (i.e., the ratio between the SNR with the amplifier on and
  the amplifier off) along the cochlea, calculated for narrowband
  near-CF signals and noise, and for broadband signals and noise
  (assumed white over the band from 4--70 kHz).  The figure shows that
  the near-CF positive SNR enhancement caused by turning on the
  amplifier produces a global, broadband increase in SNR.}
\label{fig:2}
\end{center}
\end{figure*}

{\em Preliminaries.} Figures \ref{fig:2}A,B illustrate the general
function of the mammalian ear. Briefly, sound-induced vibration of the
stapes (the third of the three middle-ear ossicles in the chain that
connects the eardrum to the cochlea) displaces the fluid in the inner
ear, launching hydromechanical waves that propagate slowly from the
base (i.e., the entrance) toward the apex (i.e., the ``end'') of the
cochlea. Cochlear wave propagation is frequency-dependent, so that
waves peak on the BM at locations that depend on frequency. In this
way, the cochlea maps frequency into position, with higher frequencies
mapping closer to the stapes. As they travel apically beyond their
peak location, cochlear waves are dramatically attenuated. Cochlear
wave propagation is also nonlinear (intensity dependent) and varies
with cochlear health (e.g., in vivo vs.\,postmortem, see
Fig.\,\ref{fig:2}B). In particular, the location of maximal vibration
depends both on sound level and on physiological status. However, at
sound levels near the threshold of hearing, where issues concerning
SNR are most pressing, cochlear mechanical responses are approximately
linear.  For this reason, we employ linear models for our analysis.
At any location we define the characteristic frequency (CF) as the
frequency that evokes the largest in vivo BM response at low sounds
levels; conversely we define the characteristic place as the location
where a wave of given frequency peaks on the BM at low sound levels.

In vivo, the cochlear amplifier boosts waves as they propagate towards
their characteristic places, producing stronger and more spatially
localized responses than in a dead cochlea
(Fig.\,\ref{fig:2}B). Equivalently, because of the well-established
symmetry between spatial and frequency tuning \cite{zweig1976}, the
cochlear amplifier narrows the bandwidth of BM frequency responses
measured at a given location (colloquially, these frequency responses
are known as ``cochlear filters''). By narrowing the bandwidth of the
cochlear filters, amplification enhances cochlear sensitivity through
well-known principles \citep{wiener1949}. Indeed, narrowing the
bandwidth of a receiver means reducing its response to background
broadband noise relative to the response to a signal within the
receiver passband. However, because it is theoretically possible to
narrow the bandwidth of the cochlear filters without resorting to
amplification \citep[e.g.\,][]{kolston1990}, we make a dedicated
effort to isolate the effects of signal amplification from the effects
of amplifier-induced bandwidth reduction.

{\em Cochlear amplification.}  In our analysis of cochlear mechanics,
we consider a general linear model that describes the frequency-domain
relationship between the velocity of the cochlear partition ($\VCP$) and the
pressure difference ($P_0$) across it. The cochlear partition comprises
the organ of Corti and the overlying tectorial membrane, and
$\VCP$ denotes the velocity of its center of mass. The
pressure-velocity relation is characterized by a phenomenological
admittance, $Y$, defined as $\VCP=YP_0$. (For simplicity,
the implicit frequency dependence is not shown.)
By applying mass conservation and Newton's second law, 
we have that (see Appendix A and
\cite{shera:nobili-waves})
\begin{equation}
\frac{1}{A}\frac{d}{dx}(A\frac{d\Pb}{dx})+\alpha ZY\Pb=0\;. \label{eq:webster}
\end{equation}
In this equation, $\Pb$ is the pressure difference between the
``upper'' and ``lower'' fluid chambers (see Fig.\,\ref{fig:2}A)
averaged over their cross-sectional area ($A$). The term $Z=i\omega M$
represents the ``longitudinal'' impedance due to the effective
acoustic mass ($M$) of the fluids, and the complex function
$\alpha=P_0/\Pb$ relates the driving pressure to the scalae-averaged
pressure \citep{duifhuis1988}; it depends on wavelength and on model
geometry. For simplicity, we assume one-dimensional (1D) wave
propagation, which allows us to set $\alpha=1$ and $\Pb=P_0$. The
equations for two- and three-dimensional (2D and 3D) models are more
complex and can be found in Appendix A. However, and as we will
illustrate through numerical simulations \footnote{Because wavenumbers
  are necessarily similar across models tailored to the experimental
  data (regardless of potentially large differences, such as their
  dimensionality \citep{zweig2015}), the results elucidated with this
  simple model are expected to apply universally to active cochlear
  models.}, the qualitative implications derived from the 1D model
remain applicable in more realistic 2D and 3D geometries.

For simplicity, our analytic treatment focuses on the amplification of
pressure, whose spatial amplification is similar to that of BM
velocity \cite{dong2013amplification}. The numerical simulations we
show in Figs.\,\ref{fig:2}D-E verify that the main results apply to BM
velocity in a more complete model \footnote{This result is not
  surprising, as pressure and velocity are related by an admittance
  function $Y$ whose value near the characteristic place is impacted
  by amplification to a much lesser extent than the pressure driving
  the motion of the BM \cite{dong2013amplification,altoe2020}.}.
Importantly, the signal enhancement mechanism we elucidate here relies
on active amplification that boosts the energy of sound-induced
traveling waves more than that of internal noise; in these types of
models, pressure amplification serves as a proxy of power
amplification \cite{altoe2020}.

When we assume ``reflectionless'' boundary conditions at the apical
and basal ends of the cochlea, the 1D Green's function becomes (see
Appendix)
\begin{equation}
G(x,x')\approx
\frac{1}{2i}\sqrt{\frac{A(x')}{A(x)}\frac{1}{{k(x)k(x')}}}\exp[
  -i\!\!\!\!\!\!\!\!
  \int\limits_{\text{min}(x,x')}^{\text{max}(x,x')}
  \!\!\!\!\!\!\!\! k(\hat x)\,d\hat x]\;, \label{eq:G1D}
\end{equation}
where $k(x)$ is the complex wavenumber.  The pressure response when
the cochlea is driven from the stapes is simply \cite{altoebox}
\begin{equation}
P(x)=2iP(0)k(0)G(x,0)\;. \label{eq:driven}
\end{equation}

\renewcommand{\Im}{\mathrm{Im}}

When the spatial gradients of cross-sectional area ($A$) and
wavenumber ($k$) are gentle enough, the gain per unit length ($g$) is
primarily determined by $\Im(k)$, the imaginary part of $k$.
Specifically, the log-gain per unit length can be approximated as
${d\log(|G|)}/{dx}\sim\Im(k)$. When $\Im(k)>0$, the gain per unit
length is greater than one, and the wave undergoes power
amplification. On the other hand, when $\Im(k)<0$, the gain per unit
length is less than one, indicating attenuation.  When the cochlear
amplifier is inactive, $\Im(k)$ is everywhere negative [$\Im(k)<0$].
But when the amplifier is maximally active, $\Im(k)$ is positive basal
to the characteristic place and negative apical to it. In other words,
the wave peaks near the point $\hat x$ where $\Im(k)=0$, with
$\Im(k)>0$ for $x<\hat x$ and $\Im(k)<0$ for $x>\hat x$ \footnote{When
  the spatial gain is sufficiently high basal to the CF place, other
  factors, such as spatial gradients of surface area and tissue
  admittance, make little contribution to determining the location of
  the wave peak.  For the arguments presented here, their
  contributions can be safely ignored.}. Importantly, waves
cut-off dramatically just apical to their characteristic place (see
Fig.\,\ref{fig:2}B), so that $g\ll1$ for $x>\hat x$. In summary,
whereas traveling waves are amplified ($g>1$) before they reach their
characteristic place ($\hat x$), they are rapidly attentuated ($g\ll1$) as they
pass beyond it.  According to our analysis of the bidirectional
amplifier [Eq.\,(\ref{eq:optimal_bi}) and Fig.\,\ref{fig:1}C], this
arrangement fulfills the conditions necessary for boosting the SNR at
the characteristic place.

{\em Amplification of narrowband signals and noise.}  For the purposes of
analyzing the effects of spatial amplification on SNR
enhancement, we focus on a narrow frequency band centered around
the signal frequency. Within an arbitrarily narrow frequency band, the
internal noise can be approximated using spatially incoherent sinusoidal
sources with randomly distributed amplitudes and phases. In particular,
we assume that the noise sources are sinusoids with phases uniformly
distributed phase on $[0,2\pi)$ and magnitudes given
  by a non-negative random variable with mean $\mu$ and variance 
  $\sigma^2$. Using this simplified noise
  model allows us to examine the impact of signal amplification on SNR
  without the confounding effects of bandwidth reduction induced by
  amplification. The rms noise pressure at a given location $x$ can be
  approximated as
\begin{equation}
\begin{aligned}
\bar{P}_\mathrm{noise}(x)\approx\gamma\sqrt{\int_0^L |G(x,x')|^2\, dx'}\;,
\end{aligned}
\end{equation}
where $\gamma^2=\mu^2+\sigma^2$. This expression represents the
statistical average of the noise pressure implied by the amplitude
distribution of incoherent sinusoidal sources. The integral
$\int_0^L |G(x,x')|^2\, dx'$ captures the propagation of noise power
from basal and apical noise sources to the location $x$.  Assuming
that the wavenumber at the cochlear entrance [$k(0)$] is independent
of cochlear amplification, we have that the SNR is \footnote{Note that
  for narrowband signal and noise, the SNR of pressure is the same as
  the SNR of velocity: the value of the partition's (or BM) admittance
  is the same for signal and noise and hence cancels out when taking
  the ratio of the two.}
\begin{equation}
R(x) \propto \frac{|G(x,0)|}{\sqrt{\int_0^x |G(x,x')|^2\, dx'+\int_x^L |G(x,x')|^2\, dx'}}\;,
\end{equation} 
where the two integrals, $\int_0^x |G(x,x')|^2 dx'$ and $\int_x^L
|G(x,x')|^2 dx'$, represent the propagated contributions of noise sources
located basal and apical to $x$, respectively.  The values
of these integrals, calculated using a previously developed 2D model (see
figure caption and Appendix B for details), are shown in
Fig.\,\ref{fig:2}C. The figure shows that at the CF place, the
contribution of apical noise sources is negligible compared to that of
basal noise sources.


Figure \ref{fig:2}D depicts the differential effects of amplification
on signal and internal noise in the 2D cochlear model for frequencies
of both 10 kHz and 30 kHz. As expected from the analysis of the
bidirectional amplifier, turning on the cochlear amplifier boosts the
signal more than the internal noise near the characteristic
place. This is evident in the plot, where the in vivo signal amplitude
is larger than that of noise near the region of maximal BM
response. (Note that signal and noise levels are normalized so that
postmortem they are the same at the characteristic place.)  However,
as one moves basally away from the characteristic place towards the
cochlear entrance, amplification becomes more pronounced for the
internal noise compared to the signal. The differential effect of
amplification on signal and internal noise highlights the selective
enhancement of the signal relative to the noise at the characteristic
place, where the cochlea achieves optimal sensitivity for sound
detection.

{\em Amplification of broadband signals and noise.} Figure
\ref{fig:2}E shows the enhancement factor as a function of distance
along the cochlea when both signals and noise are broadband. In these
simulations, signal and noise have white spectra over the frequency
band spanning the full range of CFs represented by the cochlear model
(4--70 kHz).  Except near the cochlear entrance---where CF waves do
not travel far enough to experience substantial amplification, to the
point that there is no SNR enhancement even at CF (open symbols in
Fig.\,\ref{fig:2}E)---amplification substantially boosts the broadband
SNR, by $\sim$10 dB at the most sensitive locations. These results
demonstrate that spatially restricted amplification produces a global
increase in cochlear sensitivity to broadband sounds.

\section{Discussion}

While the inner ear possesses astounding mechanical sensitivity, the
origin of this sensitivity within the context of amplification has
been largely overlooked. Indeed, the textbook view in the field is
that the cochlear amplifier increases the sensitivity of hearing by
boosting the mechanical vibrations that displace the stereocilia of
the sensory neurons. This simplistic account ignores the fact that the
sensitivity of a system depends on the internal noise
\cite{vanderheijden2021} \footnote{Indeed, the terms ``gain'' and
  ``sensitivity'' are often used interchangeably, much to the horror
  of engineers}. The handful of previous attempts at relating cochlear
amplification with (true) cochlear sensitivity
\citep[e.g.\,][]{bialek1984, martin2021} ignore the contributions of
wave propagation, relying instead on non-equilibrium oscillator models
whose relevance to cochlear mechanics remains uncertain.

We have shown here that established mechanisms of cochlear wave
amplification produce significant signal enhancement. The mechanisms
are analogous to man-made wave-based systems such as lasers and active
transmission lines \cite{chang1960, loudon1993}. Indeed, the cochlear
amplifier has been likened to the gain medium of a laser amplifier
\cite{shera2007}. By amplifying different frequencies in different
regions, the cochlea effectively employs narrowband ``laser-like''
amplification to boost sensitivity to both narrow- and broad-band
signals [Fig.\,\ref{fig:2}D]. The waveguide structure of the cochlea
allows it to act as an inhomogeneous transmission line in which the cut-off
frequency changes with location \cite{shera2005}. In this way, waves
within the operating frequency range are greatly attenuated before
reaching the apical end \citep[see also][]{puria1991,sasmal2019}.
Consequently, the cochlea eliminates noise ``build-up'' due to
scattering from the apical termination, an effect which can greatly
degrade the performance of active transmission lines \cite{chang1960}.

Our results also highlight the functional importance of the asymmetric
shape of the cochlear filters (i.e., of the BM frequency response
measured at each location).  The cochlear filters have a steep
high-frequency flank arising from the wave cut-off apical to the CF
place. As a result, near-CF waves coming from more basal locations are
amplified while those arising at more apical locations---where there
are noise sources but no signal---are squelched. Thus, the steep wave
cut-off underlies a peculiar form of spatial filtering of near-CF
components, optimized to reject noise \footnote{This spatial filtering
  mechanism is distinct from the directional amplification present in
  spatial feed-forward models \cite{steele1993, shera2023}.  Indeed,
  our analysis focuses on a model where cochlear amplifier gain is the
  same in both directions and wavenumbers are isotropic. The signal
  enhancement mechanism we deduce from this model generalizes to
  models with various degrees of anisotropy, so long as ``forward''
  waves are amplified while ``anterograde'' waves are squelched as
  they propagate towards the characteristic place.}. It is worth
noting that the ear-horn-like geometry of the cochlea contributes
significantly to this ``optimized spatial filtering.'' The tapered
geometry facilitates the propagation of waves from the base to the
apex, allowing for efficient signal propagation and amplification
\cite{altoebox}.


The strategy elucidated here for enhancing signal to noise within the
cochlea is compelling because it is simple, robust, and consistent
with established facts of active cochlear mechanics: first and
foremost, that traveling waves are initially amplified and then
dramatically attenuated as they propagate. But to what extent does
this elegant mechanism boost the sensitivity of hearing in actual
practice?  Although a precise answer to this question is currently out
of reach---it requires details that are largely unknown and are likely
to remain unknown for a long time (e.g., the power of the dominant
intracochlear noise sources)---considerable insight can be gained by
reviewing the empirical evidence in light of our
findings. Specifically, Nuttall and colleagues \cite{nuttall1997}
measured BM-velocity noise in the base of sensitive guinea-pig
cochleae, carefully minimizing external interferance to ensure that the
recordings were dominated by internal cochlear noise sources.  At
frequencies near CF, they found a BM mechanical noise floor
approximately 15 dB below the BM vibration amplitude produced by tones
at intensities corresponding to neural threshold. More recent
recordings \cite{quinones2022}, in the apex of the mouse cochlea,
yield similar results for the tectorial membrane \footnote{Near CF,
  the tectorial membrane noise floor in the $\sim8$-kHz region of the
  healthy mouse cochlea ($0.1\,$nm or smaller) is roughly 10 to 15 dB
  smaller than the vibration evoked by 10 dB SPL tones at CF
  ($0.3-0.4\,$nm) \cite{quinones2022}. In this region of the mouse
  cochlea, the threshold of the most sensitive auditory-nerve fibers
  is on the order of 10 dB SPL \cite{taberner2005}.}. In a nutshell,
the experimental data suggest that the cochlear mechanical SNR,
measured for narrowband frequencies near-CF in response to
threshold-level tones, is on the order of 15 dB. Strikingly, in our
model amplification enhances the SNR of the BM responses by a similar
amount (Fig.\,\ref{fig:2}E). In other words, our results suggest that
without amplification cochlear mechanical responses to faint but
detectable sounds would fall perilously close to the internal noise
floor. Although there is no scarcity of factors that impact the neural
encoding of sound---including hair-cell noise
\cite{holton1986,beurg2021} and the stochastic nature of
auditory-nerve firing \cite{siebert1970}---our analysis suggests that
spatially distributed cochlear amplification plays a central role in
enhancing the sensitivity of hearing.

\section*{Acknowledgments}
Supported by grants R21 DC019712 (AA) and R01 DC003687 (CAS) from the
NIDCD/NIH.

\appendix
\section{Green's Functions in 1, 2, and 3 Dimensions}
{\em Equations of motion.}  The average pressure difference between
the two scalae ($\Pb$) and the velocity of the partition's
center of mass 
($\VCP$) are related by the well-known transmission-line equations:
\begin{equation}
\begin{aligned}
\dx{\Pb}&=-\frac{i\rho\omega}{A}U,& \label{eq:tl1} \\
\dx{U}&=-b\VCP\;,& 
\end{aligned}
\end{equation}
where $U$ is the volume velocity of the fluids in the duct, $A$
is the duct's effective acoustic area, $\rho$ is the fluid density,
$\omega$ is the angular frequency, and $b$ is the partition's effective width.
The partition velocity can be expressed as the product of the pressure difference across the tissue $P_0$ and a complex admittance $\YCP$
\begin{equation}
\VCP=P_0\YCP=\alpha\Pb\YCP \label{eq:vy}\;,
\end{equation}
where $\alpha=P_00/\Pb$ is the short-wave hydrodynamic factor. Combining Eqs.\,(\ref{eq:tl1}--\ref{eq:vy}) we find the following expression for $\Pb$
\begin{equation}
\frac{1}{A}\frac{d}{dx}(A\dx{\Pb})+k^2_x\Pb=0,
\end{equation}
where $k_x^2=\alpha Z Y$ is the square of the complex wavenumber
and $Z={i\rho\omega}/{A}$
is the acoustic impedance of the scalae.

{\em 1D models.}  In 1D models, the pressure field is a function only
of longitudinal distance from the stapes ($x$) so that $\Pb\equiv P_0$
and $\alpha=1$. The Green's function $\GD(x,x')$ is the response to a
unitary point pressure source at $x'$, and hence can be expressed as
\begin{equation}
\frac{1}{A}\frac{d}{dx}(A\dx{\GD})+k^2_x{\GD}=-\delta(x-x')\;, \label{eq:horn}
\end{equation}
where $\delta$ is the Dirac delta function.
Note that the pressure source has unit of pressure over length
squared. We assume reflections boundary conditions and calculate $\GD$
using the Wentzel-Kramers-Brillouin (WKB) approximation.  To do so we
make a change of variable in Eq.\,(\ref{eq:horn}). In particular, we
introduce the acoustic distance $\chi$,
\begin{equation}
\chi=A(0)\int_0^x\frac{dx}{A(x)}\;,
\end{equation}
and define the corresponding wavenumber as $\hat k=A
k_x/{A(0)}$. Equation (\ref{eq:horn}) can be then rewritten as
\begin{equation}
\begin{aligned}
\dxii{\GD}+\hat k^2\GD=-\frac{A^2}{A^2(0)}\delta(x-x')=  \label{eq:imp_horn}\\
=-\frac{A(x')}{A(0)}\delta(\chi-\chi')\;. 
\end{aligned}
\end{equation}
In the $\chi$ domain, the 1D Green's function is \cite{shera2005}
\begin{equation}
  G(\chi,\chi')=\frac{1}{2i}\sqrt{\frac{1}{\hat k(\chi) \hat k(\chi')}}
  \exp[-i
    \!\!\!\!\!\!\!
    \int\limits_{\min(\chi,\chi')}^{\max(\chi,\chi')}
    \!\!\!\!\!\!\!
    \hat k(\hat\chi)d\hat\chi].
\end{equation}
Accounting for the source amplitude in the $\chi$ domain
[Eq.\,(\ref{eq:imp_horn})],
and converting the solution back into the $x$ domain, yields
 \begin{equation}
\GD(x,x')= 
\frac{1}{2i}\sqrt{\frac{A(x')}{A(x)}\frac{1}{k(x)k(x')}}\exp[-i
    \!\!\!\!\!\!\!
  \int\limits_{\min(x,x')}^{\max(x,x')}
\!\!\!\!\!\!\!\!
             {k(\hat x)d\hat x}]\;.
       \label{eq:G1D}
\end{equation}

{\em 2D and simplified 3D models.}  A tapered 2D ``box'' model can be
interpreted physically as a model where the cross-sectional area of
the duct is a rectangle with constant width and varying height, while
the partition spans the entire cochlear width and moves up and down as
a piston (``wall-to-wall carpeting'', see \cite{shera2005}).  The
equations for a 2D model are approximately valid for a 3D model where
the cochlear duct and partition have circular cross-sectional shapes.
When the radius of the partition is sufficiently small, the pressure
can be approximated as a function of distance from the stapes ($x$)
and radial distance from the partition center ($r$). With these
approximations the 3D model is effectively 2D in cylindrical
coordinates \citep{deboer1981, altoe2022}.

Importantly, although the equations for a
2D box model are valid for a 3D
cylindrical model, the parameters and their spatial gradients in
the two models are diffferent \citep[see also][]{altoe2020}. While
the partition admittance can be strategically chosen so that
the wavenumber $k_x$ is the same in 2D
and 3D \footnote{If the wavenumber $k_x$ is identical in the two models,
  $\alpha$ is also expected to be the same because its WKB approximation is
  the same [${\alpha_\mathrm{WKB}\approx kH/\tanh(kH)}$].}, the spatial
gradient of the cross-sectional area $A$ (which determines the
important geometric pressure gain factor \cite{altoebox}) differs in
the two models. In the tapered box model $A\propto H$, while in the 3D
cylindrical model $A\propto H^2$, where $H$ is the scala height (or radius).

Keeping in mind these important caveats, we now proceed to
heuristically determine the reduced 2D Green's function $\GDb(x,x')$
which describes the scalae-average pressure at $x$
resulting from a 2D source
placed at the center of the partition (i.e., at $y=0$).  Following
\cite{shera2005}, we note that the 2D reduced Green's function must
obey the following relation
\begin{equation}
\begin{aligned}
&\frac{1}{A}\frac{d}{dx}(A\dx{\GDb(x,x')})+k^2_x{\GDb(x,x')}=\mathcal{F} \delta(x-x')\;,& \label{eq:horn3D} \\
\end{aligned}
\end{equation}
where $\mathcal{F}$ is a function to be determined that accounts for
the fact that the source, unlike in the 1D model, is also
two-dimensional. Following the results of \cite{shera2005} obtained in
a box model of constant cross-sectional area, we have that
$\mathcal{F}|_{x'}\propto\alpha(x')$. Because in our tapered model the
area changes with location, we further need to figure out if there are
systematic differences between 1, 2, and 3D sources that change with
the cross-sectional area. In this regard, we note that a 2D point
source is $s_\mathrm{2D}=\delta(x-x')\delta(y)$ while a
one-dimensional source is $s_\mathrm{1D}=\delta(x-x')$. Their
respective source strengths, averaged over the cross sectional area of
a two-dimensional model, are a factor of $H(x')$ larger in 1D than in
2D \footnote{The source strength, averaged over the cross sectional
  area of the cochlea, is $\frac{1}{A}\iint_{A}s dS$.}. Likewise a 3D
source is $s_\mathrm{3D}=\delta(x-x')\delta(y)\delta(z)$, whose
strength is $A(x')$ smaller than a 1D one.

Based on these consideration, and further noting that $A(x')\propto
H(x')$ in 2D (so that we can write equations that are valid in 2D and
3D models), we conclude that $\mathcal{F}\approx \alpha(x')/A(x')$:
\begin{equation}
\GDb(x,x')\approx\frac{\alpha(x')}{A(x')}\GD(x,x'). \label{eq:G12D}
\end{equation}
We now define $\GDD(x,x')=\GDDD(x,0,x',0)$ where $\GDDD(x,y,x',y_0)$
is the ``true'' 2D Green's function, i.e., the function that describes
the pressure response at  $x,y$ to a unit point source at
$x',y'$. Exploiting the definition of $\alpha$, we have that
 \begin{equation}
 \begin{aligned}
\GDD(x,x')&\approx\\
\frac{\alpha(x)\alpha(x')}{2i}&\sqrt{\frac{1}{k(x)k(x')A(x')A(x)}}\exp[-i
    \!\!\!\!\!\!\!
  \int\limits_{\min(x,x')}^{\max(x,x')}
    \!\!\!\!\!\!\!
             {k(\hat x)d\hat x}], \label{eq:G2D}
 \end{aligned}
\end{equation}
where it can be appreciated that $\GDD(x,x')=\GDD(x',x)$. The same
Green's function holds for the simple 3D model
[${\GTD(x,x')\approx\GDD(x,x')}$], keeping in mind the caveats
regarding the cross-sectional areas in the two models.

{\em Non-ideal boundary conditions and numerical solutions.}  The
solutions for the Green's functions shown above were obtained under the
assumption of no significant scattering from the basal and
apical boundaries. While this is a good approximation for the apical
boundary---because traveling waves are dramatically attenuated before reaching
it \cite{puria1991}---the same is not true for the basal boundary at the stapes,
where any impedance mismatch at the boundary with the middle ear
has the effect of backscattering a significant fraction of wave power
\cite{shera1991}. When the wave frequency is sufficiently smaller than
the CF near the stapes we can assume long-wave behavior near the
stapes. In this case, we can easily include
the effect of wave reflection
and calculate the WKB approximation for this non-idealized Green's
function
\begin{equation}
\begin{aligned}
\GDN(x,x')= & \;\GDD(x,x') \; +\\
+R_\mathrm{st}\GDb(0,x')&\alpha(x)\sqrt{\frac{A(0)k(0)}{A(x)k(x)}}\exp{-i\int\limits_0^xk(\hat x)d\hat x}, \label{eq:gnonid}
\end{aligned}
\end{equation}
where $R_\mathrm{st}$ is the complex reflectance of the stapes
\cite{shera1991}. The second term on the right side of
Eq.\,(\ref{eq:gnonid}) represents a wave traveling from the base to the
apex, generated by the pressure reflected from the stapes
[$R_\mathrm{st}\GDb(0,x')$].

\begin{figure}[htbp]
\begin{center}
\includegraphics{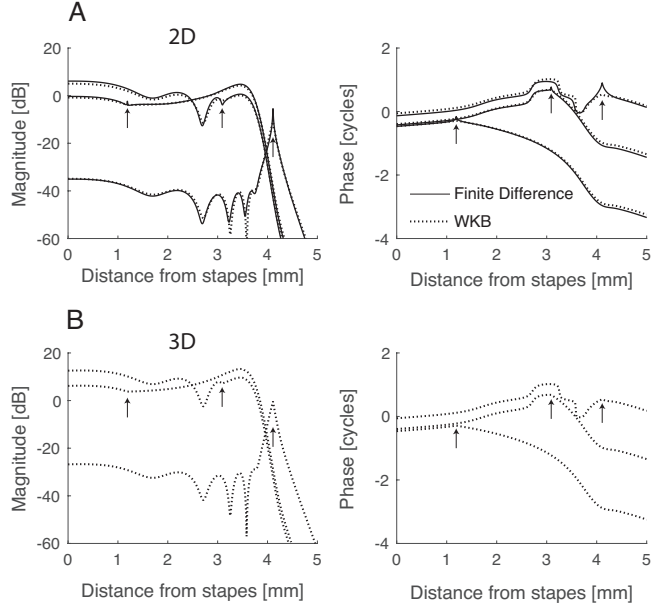}
\caption{ A) Example of Green's function for a 2D model with
  reflective basal boundary ($|R_\mathrm{st}|\approx0.14$), calculated
  numerically in a finite-difference model (solid line) or with the
  WKB approximation [Eqs.\,(\ref{eq:G2D},\ref{eq:gnonid}), dashed
    lines]. The source locations for the various curves are indicated
  with vertical arrows; the source frequency is 10 kHz. B) Approximate
  Green's function for a simplified 3D model (see text).}
\label{fig:appendixA}
\end{center}
\end{figure}

{\em Numerical and semi-analytical calculations} We cross-checked the
quality of our calculations by comparing the 2D WKB approximation of the
Green's function against numerical calculations performed in a tapered
2D finite-difference model \cite{neely1981,altoe2023}, some of
which are shown in Fig.\,\ref{fig:appendixA}A. Because calculating the
WKB approximation for $\alpha$ requires iterative methods that
introduce various inaccuracies, we calculated $\alpha$ numerically,
driving the finite-difference model from the stapes.  Figure
\,\ref{fig:appendixA}B shows the WKB solution for the Green's function
of a 3D model with the same wavenumber ($k$) and height ($H$) as the 2D
model in Fig.\,\ref{fig:appendixA}A. While the agreement between the WKB
approximation and the numerical solution is generally excellent, the
WKB approximation can introduce significant errors (due to
the non-uniqueness of the WKB solution in the cut-off region
\citep{watts2000}), rendering the calculations noisy, especially at high
frequencies. For this reason, in the main text we present results
obtained using the 2D finite-difference model---the differences between
2D and 3D models are relatively minor, although it is worth mentioning
that in 3D the enhancement factors are slightly larger thanks to the
more dramatic tapering of the cross-sectional area in 3D than in 2D
models.

\section{Modeling Details}
We performed all calculations using an ``overturned model'' of the
mouse cochlea \cite{altoe2022}, whose parameters are the same as those
used in \cite{shera2023}. In this model, unlike in classic models
where the organ of Corti does not deform, the transverse (up-down)
velocity of the center of mass is $\VCP=(\VBM+\Vtop)/2$, where $\VBM$
and $\Vtop$ indicate the velocity of the bottom (BM) and the top-side
(the reticular lamina and tectorial membrane) of the organ of
Corti---their differential velocity is $\Vint=\Vtop-\VBM$.  Postmortem,
$\VBM$ and $\Vtop$ are similar, so that to a first approximation
$\Vint\approx0$ in a passive cochlea, while $\Vint\neq0$ in vivo. The
center-of-mass velocity can be rewritten in the compact form
$\VCP=\VBM+\Vint/2$, where $\Vint$ is attributed to the piezo-electric
action of the OHCs and is effectively the (velocity)
source of wave amplification in the model.
 
Because the BM stiffness is about one order of magnitude larger than
that of the structures surrounding the OHCs, OHC forces produce large
displacements of the top side of the organ of Corti while having
secondary effects on local BM motion \cite{dewey2019,dewey2021}. We
therefore assume that internal OHC forces have negligible effects on
BM motion so that the mechanical admittance of the BM
($\YBM=\VBM/P_0$) is constant, independent of whether the cochlear
amplifier is turned ``on'' or ``off''. For simplicity, we also assume
that $\YBM$ represents the admittance of a damped harmonic oscillator.
By exploiting the relationships between $\VCP$, $\VBM$, and $\Vint$,
we can express the admittance of the organ of Corti admittance as
$\YCP=\YBM(1+\frac{1}{2}\Vint/\VBM)$. Following previous results, we
assume that in vivo at low sound levels $\frac{1}{2}\Vint/\VBM\approx
i\beta\tau$, where $\beta=f/\mathrm{CF}$ is normalized frequency and
$\tau$ is a (real) constant. Following \cite{altoebox}, we assume that
$\YCP$ is scaling symmetric (i.e., a function only of normalized
frequency, $\beta$) throughout the cochlea.

\bibliography{bib_noise}

\end{document}